\documentclass[traditabstract]{aa}
\usepackage[utf8]{inputenc}
\usepackage{amsmath}
\usepackage{natbib}
\usepackage{graphicx}
\usepackage{pdfsync}
\usepackage{txfonts}
\usepackage{color}
\usepackage{multirow}
\usepackage{amssymb}
\usepackage[plainpages=false]{hyperref}
\usepackage{xspace}

\title{Her X-1: the positive cyclotron line energy / luminosity correlation}
\author{Davide Vasco\inst{1}, Dmitry Klochkov\inst{1}, and R\"udiger Staubert\inst{1}}
\institute{Institut f\"ur Astronomie und Astrophysik, Sand 1, 72076 T\"ubingen, Germany}

\begin{document}

\abstract{Studies of some bright, super-Eddington transient pulsars show a negative correlation between 
the energy of the cyclotron resonance scattering feature (CRSF) and the bolometric luminosity. For 
Her~X-1, using  repeated \textsl{RXTE} observations during 1996--2005, the inverse dependence was found:
the energy of the cyclotron line increases as the luminosity increases. The X-ray flux measured
by the \textsl{RXTE}/ASM (2--10\,keV) has been assumed to represent the luminosity - more precisely: 
the maximum X-ray flux reached during the respective 35\,d \textsl{Main-On}. Here, we question 
whether the ASM flux is really an accurate measure of the bolometric luminosity of the source. 
We redetermined the energy of the cyclotron line and performed spectral fits using the combined data from 
the PCA (3.5--60\,keV) and HEXTE (20--75\,keV) instruments on \textsl{RXTE} of the same 35\,d cycles as used 
in the original work to determine the bolometric flux from those spectra. We confirm the result of the 
original analysis that the cyclotron line energy changes by $\sim$7\% for a change in flux by a factor of two. }

\keywords{pulsars: individual: – stars: neutron – stars: binaries}
\authorrunning{D. Vasco et al.}
\maketitle

\section{Introduction}

Her X-1 is one of the most observed and well-studied accreting binary X-ray pulsars. Since the first observation 
made by  \textsl{Uhuru}, the 35\,d periodicity of the source is well known. The X-ray light curve shows two 
{\em on}-states (high X-ray flux) and two {\em off}-states (low X-ray flux), with  a \textsl{Main-On} 
($\sim$ 7 orbital periods) and a \textsl{Short-On} ($\sim$ 5 orbital cycles), separated by two  
{\em off}-states ($\sim$ $4\div5$ orbital cycles each). The maximum X-ray flux of the \textsl{Main-on} is higher 
than the maximum X-ray flux of the \textsl{Short-On} by a factor of three to four.

The modulation of the X-ray flux is believed to be caused by the periodic obscuration of the X-ray source by the precessing 
disk, which is believed to be inclined and warped \citep{Gerend_Boynton_76}. The onset of the flux, the \textsl{turn-on}
(often identified with 35\,d phase 0.0), is believed to occur when the outer rim of the disk opens up the view
to the X-ray emitting regions near the polar caps on the surface of the neutron star, while the flux decrease
towards the end of the {\em on}-states is associated with the inner parts of the disk covering these regions from the 
observer.

Her X-1 is also the first X-ray pulsar  for which a cyclotron line was discovered \citep{Truemper_etal_78}. 
The discovery of this line-like feature played a key-role in the measurement of the magnetic fields, providing
the first direct measurement of the B-field of a neutron star. The energy of the cyclotron line is related to
the magnetic field by the formula B$_{12}$ = (1+z) E$_{\rm cyc}$/11.6\,keV, where B$_{12}$ is the 
magnetic field strength in units of $10^{12}$\,gauss, $z$ is the gravitational redshift and E$_{\rm cyc}$ is the 
energy of the cyclotron line. This feature is now referred to as a cyclotron resonant scattering feature (CRSF) 
and seems to be quite common in accreting X-ray pulsars 
\citep{Coburn_etal_02}. The cyclotron line is an absorption feature produced by 
the resonant scattering of photons on electrons. In the $\sim10^{12}$\,gauss magnetic field, the
electrons are in quantized energy states (with respect to their movement perpendicular to the magnetic field),
the so-called Landau levels. The energy levels are nearly equidistantly spaced and photons with energies equal 
to $n$ times the fundamental Landau energy could take part in this scattering.

In a few transient pulsars such as V0332+53 and 4U 0115+63, a negative correlation between the CRSF 
and the bolometric luminosity of the source has been observed:  the energy of the cyclotron line decreases as the X-ray luminosity increases \citep{Mihara_etal_98,Mowlavi_etal_06,Nakajima_etal_06,Tsygankov_etal_06}. 
Her X-1, however, shows the opposite behavior. Repeated measurements of the cyclotron line energy with 
different instruments such as \textsl{RXTE} and \textsl{INTEGRAL} in the X-ray spectrum of Her X-1 have 
revealed  a \textsl{positive} correlation between the (pulse phase-averaged) cyclotron line energy E$_{\rm cyc}$ and 
the X-ray flux \citep{Staubert_etal_07}, more precisely with the \textsl{maximum X-ray flux} observed during the 
\textsl{Main-On} of the respective 35\,d cycle, as measured by \textsl{RXTE}/ASM in the 2--10\,keV range, which is
assumed to be characteristic of the current accretion state and luminosity of the source.
It has, however, been questioned whether the 2--10\,keV flux can really be taken as representative of the
bolometric luminosity.

In this paper, we attempt to answer this question by re-analysing the \textsl{RXTE} observations of the 
same 35\,d cycles as used in the original analysis of \citet{Staubert_etal_07}. Using data from both instruments 
(PCA and HEXTE), we re-determine the cyclotron line energy and measure the 3.5--60\,keV bolometric X-ray flux 
by performing spectral analyses of eleven \textsl{Main-Ons} between 1996 and 2005. 
We show that the 2--10\,keV flux is a good measure of the bolometric flux and that the positive correlation 
between the phase-averaged cyclotron line energy E$_{\rm cyc}$ and the \textsl{maximum 35\,d flux}, 
or the luminosity, is confirmed. 

%------------ TABLE 1 -----------------
\begin{table}
\begin{center}
\vspace*{1mm}
\caption{Details of \textsl{RXTE} observations of Her~X-1 used for the spectral analysis.} 
\label{obs_table}
\resizebox{0.38\textwidth}{!}{
\begin{tabular}{l c c }
\hline\hline
Observation & 35~d Main-On              & Center  \\
month/year  & cycle number  $^{1}$    & MJD \\
  \hline
July 96             &  257  & 50029.75 \\
September 97  &  269  & 50707.06 \\ 
December 00   &  303  & 51897.69 \\
January 01       &  304  & 51933.67 \\
May 01             &  307  & 52035.48 \\
June 01            &  308  & 52071.16 \\
August 02         &  320  & 52492.96 \\
November 02   &  323   & 52599.32 \\
December 02   &  324   & 52634.01\\
October 04       &  343  & 53300.95 \\
July 05              &  351  & 53577.35 \\
\hline
\end{tabular} }
\vspace*{1mm} \\
$^{1}$    Cycle numbering according to \citet{Staubert_etal_09}
\end{center}
\label{table1}
\end{table}
%------------------------------------

\vspace*{-5mm} 
\section{Observations} 
Her X-1 has been repeatedly observed by \textsl{RXTE} since 1996. We analyzed observations 
of eleven 35\,d \textsl{Main-Ons} for which there was photon statistics of sufficient high quality to allow a  spectral analysis. 
Those cycles were those of numbers (nos.) 257, 269, 303, 304, 307, 308, 320, 323, 324, 343 and 351. For the definition of cycle
counting we refer to \citet{Staubert_etal_09} (cycle no. 313 was not included, despite the high quality statistics, because no
ASM observations were available). We cover a period of ten years of observation from July 1996 (no. 257)
to July 2005 (no. 351), see Table~\ref{table1}. Here we used data from both \textsl{RXTE} instruments: PCA
in the energy range  3.5--60\,keV, and HEXTE in the energy range  20--75\,keV.

\section{Spectral analysis} 

For the spectral analysis, we used XSPEC\footnote{http://heasarc.gsfc.nasa.gov/docs/xanadu/xspec}
 (12.6.0) and generated spectra for all observed \textsl{Main-Ons}, summing all available data. 
For each spectrum (one for each  \textsl{Main-On}), two quantities were determined: the central energy 
E$_{\rm cyc}$ of the cyclotron absorption feature and the integrated flux in the range 3.5--60\,keV 
(in units of $\rm erg~cm^{-2}~s^{-1}$). 
As the spectral function, we used the \texttt{highecut}$^1$ model (based on a power-law continuum with an exponential 
cut-off) and a multiplicative \textsl{Gaussian} absorption line for the cyclotron resonant scattering feature 
(CRSF), in the same way as e.g. \citet{Coburn_etal_02} and \citet{Staubert_etal_07}. Data from both   
\textsl{RXTE} instruments were used: PCA (PCU2 only) in the energy range  3.5--60\,keV, and HEXTE in the 
energy range  20--75\,keV.  
Cold material absorption was taken into account in each fit. All values were consistent to within two standard
deviations with the mean value of $N_{H}$ = $1.08\times10^{22}$\,cm$^{-2}$.
We note that in the original analysis PCA data were used only up to 25\,keV to define the continuum at the lower energies. 
New response matrices are now available\footnote{http://www.universe.nasa.gov/xrays/programs/rxte/pca/doc/rmf/
pcarmf-11.7} that allow us to use the PCA up to $\sim$50\,keV, but
we found that the PCA can indeed be used up to 60 keV,
thereby contributing information about the cyclotron line around 40\,keV. This choice is confirmed by the agreement between the  PCA and HEXTE spectra in the overlapping region of the two instruments (40--60\,keV). This is also confirmed in the analysis  by \citet{Rothschild_etal_11} of RXTE observations of Cen A in which PCA data were successfully used up to 60 keV.
In the new spectral fits, we also added systematic uncertainties of 0.5\% (which are recommended for use with the new
response matrices$^2$), while in the original analysis by \citet{Staubert_etal_07} 1\% uncertainty was added.

For this spectral re-analysis, only data of \textsl{RXTE} PCU2 were used to ensure maximum 
uniformity in the data set, analyzed in a uniform way. Fig. 1 shows an example of 
a spectral fit to data of July 1996 (cycle no. 257). The observation is centered at MJD~52599.36
and the total integration time is 31\,ks for PCA/PCU2 and 10\,ks for each of the two HEXTE clusters. 

% Fig. 1
%\begin{figure} [b!]
\begin{figure}
\begin{center}
\includegraphics[bb=41 17 739 760, angle=-90, width=9cm]{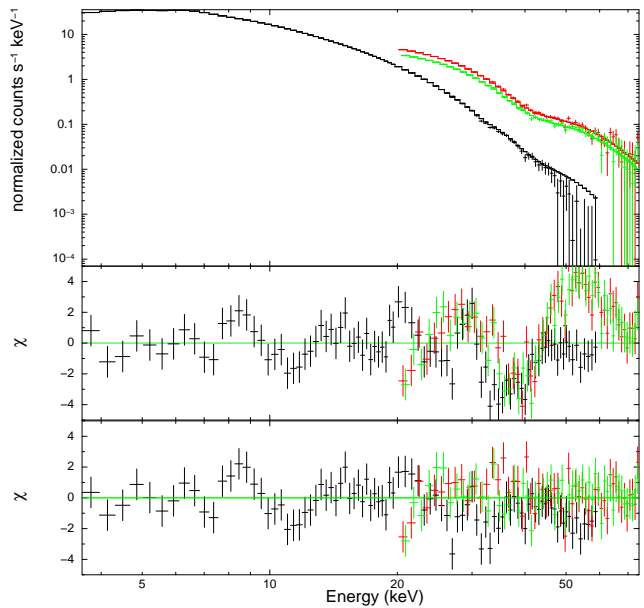} 
\vspace{2mm}
\caption{Example of a spectral fit to an \textsl{RXTE} observation of Her~X-1 in July 1996 (cycle no. 257). The observation is centered at MJD~52599.36
and the integration time is 31\,ks for PCA and 10\,ks for each of the two HEXTE clusters. Black: PCA, red: HEXTE-A, and green: HEXTE-B, respectively;  \textsl{top}: count rate spectra; \textsl{middle}: residuals with respect to a  fit of a continuum model; \textsl{bottom}: residuals with respect to a fit which includes a cyclotron line.} 
\end{center}
\label{spec} 
\end{figure}

The bolometric flux for the individual spectra was found by integrating the fit function (over the 3.5--60\,keV range),
using the XSPEC routine \texttt{flux}$^1$. The X-ray flux of Her~X-1 varies as a function of phase of the 35\,d 
modulation because of variable absorption by the accretion disk.

% Fig. 2
\begin{figure*}[t]
     \includegraphics[width=0.50\textwidth]{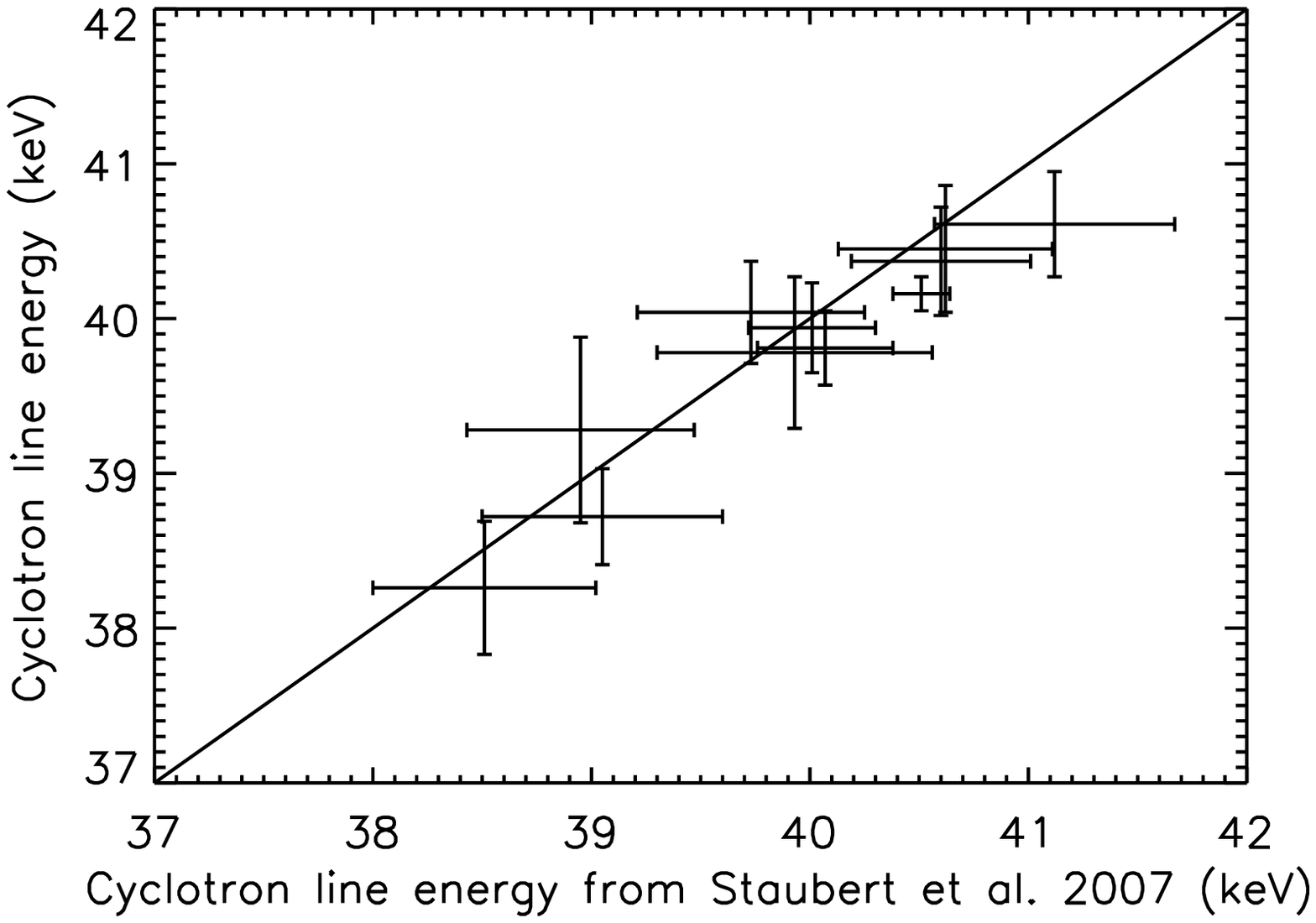}
     \hfill
     \includegraphics[width=0.50\textwidth]{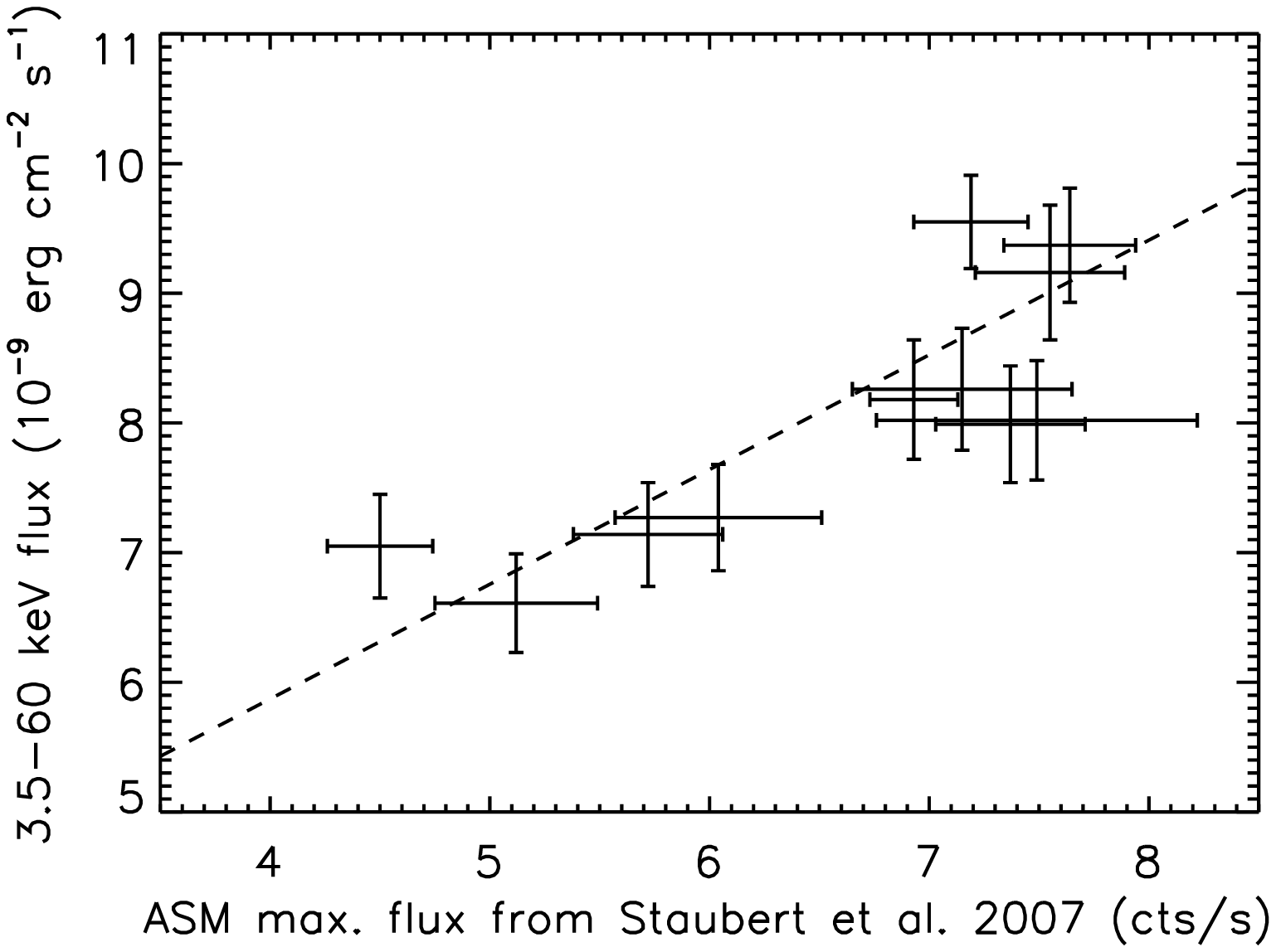}
   \caption{\textsl{Left}: E$_{\rm cyc}$ from the spectral re-analysis versus E$_{\rm cyc}$ from the original 
   analysis \citep{Staubert_etal_07}. 
   \textsl{Right}: Scaled maximum 3.5--60\,keV flux (in units of $\rm 10^{-9}~erg~cm^{-2}~s^{-1}$) from the spectral 
   re-analysis versus the maximum 2--10\,keV ASM flux (in units of $\rm cts~s^{-1}$) from the original analysis. 
   The Pearson correlation coefficients are 0.95 (left) and 0.82 (right).}
   \label{compar}
\end{figure*}

%------------ TABLE 2  -----------------
\begin{table*}[t!]
\begin{center}
\caption{Phase-averaged cyclotron line energies E$_{\rm cyc}$ and \textsl{maximum fluxes} 
reproduced from \citet{Staubert_etal_07}~(1) and from the spectral re-analysis of this work (2).}
\begin{tabular}{lcccc}
\hline \hline
 35\,d cycle      & E$_{\rm cyc}$(1) [keV]    & \textsl{max. flux} (2--10\,keV) & E$_{\rm cyc}$(2) [keV] & \textsl{max. flux} (3.5--60\,keV) $^{2}$\\
 number$^{1}$ &                                         & ASM [$\rm cts~s^{-1}$]          &                                      & [$\rm 10^{-9}~erg~cm^{-2}~s^{-1}$] \\
                        & (\citet{Staubert_etal_07}) & (\citet{Staubert_etal_07})       & (this work)                    & (this work) \\
\hline
 257 & $41.12 \pm 0.55$   &  $7.37 \pm 0.34$   &  $40.61 \pm 0.34$  &  $ 7.99 \pm 0.45$ \\
 269 & $40.62 \pm 0.49$   &  $7.49 \pm 0.73$   &  $40.45 \pm 0.41$  &  $ 8.02 \pm 0.46$ \\ 
 303 & $40.07 \pm 0.31$   &  $6.04 \pm 0.47$   &  $39.81 \pm 0.24$  &  $ 7.27 \pm 0.41$ \\
 304 & $39.05 \pm 0.55$   &  $5.72 \pm 0.34$   &  $38.72 \pm 0.31$  &  $ 7.14 \pm 0.40$ \\
 307 & $39.93 \pm 0.63$   &  $7.15 \pm 0.50$   &  $39.78 \pm 0.49$  &  $ 8.26 \pm 0.47$ \\
 308 & $39.73 \pm 0.52$   &  $6.93 \pm 0.20$   &  $40.04 \pm 0.33$  &  $ 8.18 \pm 0.46$ \\
 320 & $40.01 \pm 0.29$   &  $7.19 \pm 0.26$   &  $39.94 \pm 0.29$  &  $ 9.55 \pm 0.36$ \\
 323 & $40.51 \pm 0.13$   &  $7.64 \pm 0.30$   &  $40.16 \pm 0.11$  &  $ 9.37 \pm 0.44$ \\
 324 & $40.60 \pm 0.41$   &  $7.55 \pm 0.34$   &  $40.37 \pm 0.35$  &  $ 9.16 \pm 0.52$ \\
 343 & $38.51 \pm 0.51$   &  $4.50 \pm 0.24$   &  $38.26 \pm 0.43$  &  $ 7.05 \pm 0.40$ \\
 351 & $38.95 \pm 0.52$   &  $5.12 \pm 0.37$   &  $39.28 \pm 0.60$  &  $ 6.61 \pm 0.38$ \\
\hline 
\end{tabular} 
\vspace*{-1mm} \\
\end{center}
$^{1}$  Note that for 35\,d cycle numbers 303 and larger the corresponding numbers 
in \citet{Staubert_etal_07} are larger by 1. This reflects the observation that there must have been an extra 
cycle during the long \textsl{anomalous low} before cycle 303. However, using the numbers as given here 
allows to use them in an ephemeris for a rough prediction of the 35\,d turn-ons using a mean period of 35.88\,d. 
We do not doubt the physical reality of the extra cycle found by \citet{Staubert_etal_07}. \\
$^{2}$ The flux in Col. 5 represents the 35\,d \textsl{maximum flux} in the 3.5--60\,keV range (see text).
\label{table2} 
\end{table*}
%-----------------------------

The sampling of the various \textsl{Main-Ons} in groups of pointed observations was rather different, and the
fluxes determined from the spectra are mean values of the observations, which happened to be made at different 
phases of the 35\,d modulation.These fluxes cannot directly be compared to the  \textsl{maximum X-ray flux} of the 
respective \textsl{Main-On}. To find the comparable bolometric fluxes in the 3.5--60\,keV range that  
represent the \textsl{maximum flux} for the particular cycle, it was necessary to scale the fluxes found in the individual 
spectra to the \textsl{maximum flux}. The scaling factor is  the ratio of the \textsl{maximum flux} to the mean 
flux of the complete light curve, that is of the same data used to generate the spectra.
To find the \textsl{maximum flux}, all observed \textsl{RXTE}/PCA light curves were fitted by a function representing a 
template of the mean 35\,d \textsl{Main-On} modulation, the shape of which was taken from an overlay of many 35\,d light 
curves observed by \textsl{RXTE}/ASM (see e.g. \citealt{Klochkov_etal_06}). For this procedure, we are only 
interested in the overall 35\,d modulation, that is without eclipses and dips, which was described by the  
analytical function: 
$$
f(t) =
A_0\left( 1-\frac{A_1}{1+\exp{(B_1({t-C_1))}}}\right)\frac{1}{1+\exp{(B_2(t-C_2))}},
$$

% Fig. 3
\begin{figure*}[t]
     \includegraphics[width=0.50\textwidth]{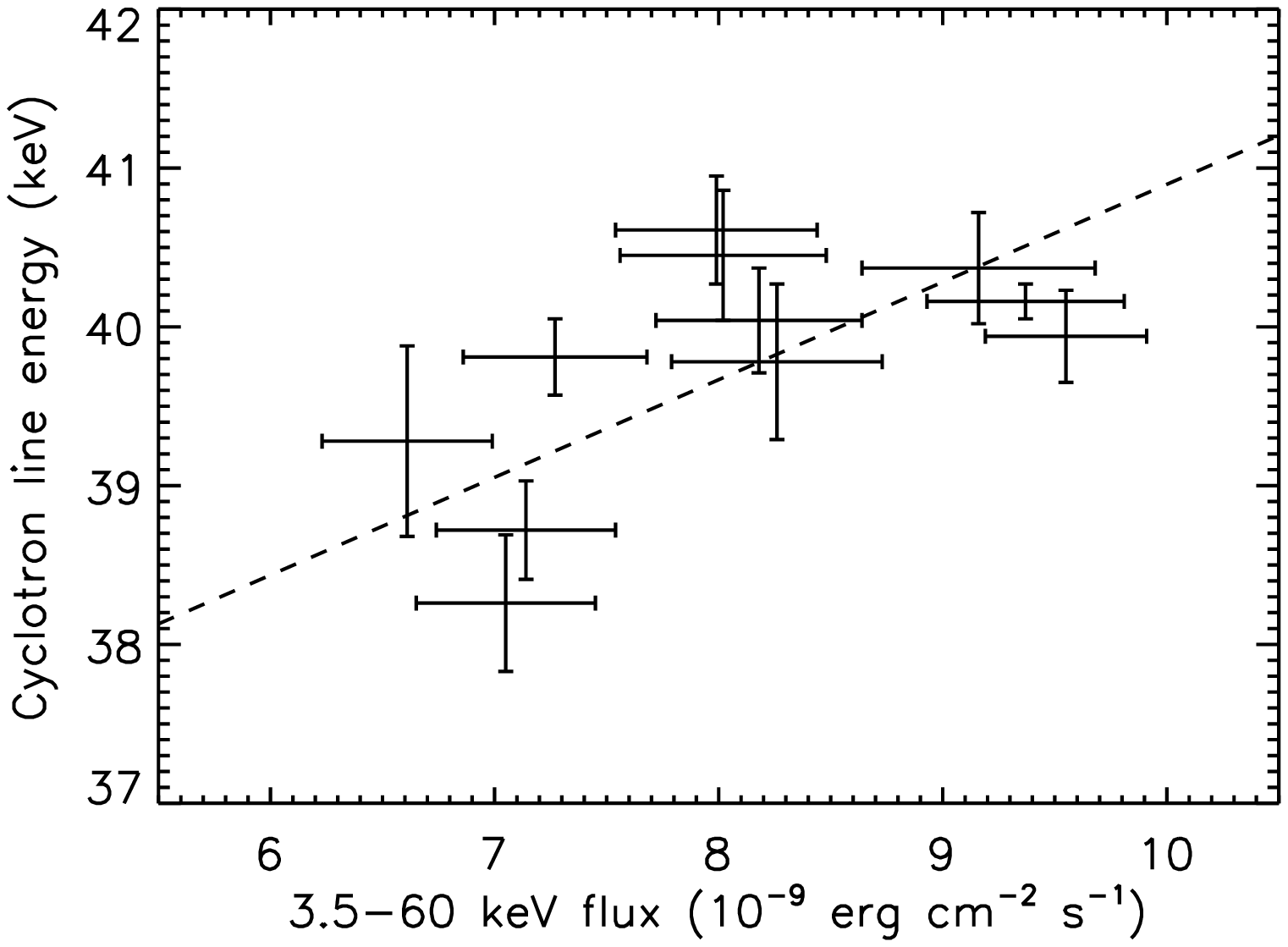}
     \hfill
     \includegraphics[width=0.50\textwidth]{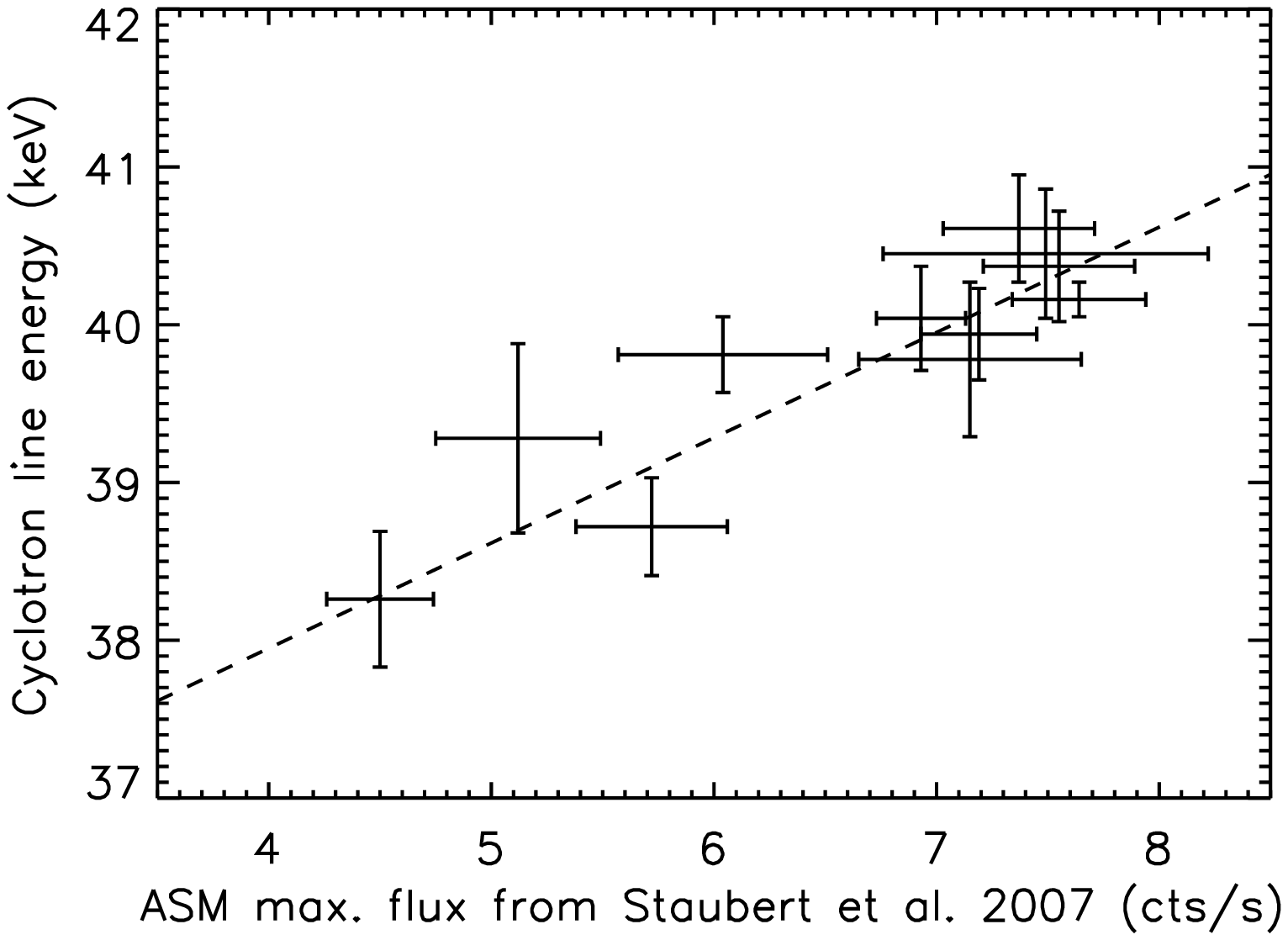}
   \caption{\textsl{Left}: E$_{\rm cyc}$ from the spectral re-analysis versus the scaled maximum 3.5--60\,keV flux 
   (in units of $\rm 10^{-9}~erg~cm^{-2}~s^{-1}$). \textsl{Right}: E$_{\rm cyc}$ from the spectral re-analysis versus the 
   maximum 2--10\,keV ASM flux (in units of $\rm cts~s^{-1}$) from the original analysis. }
   \label{correl}
\end{figure*}

\noindent under the condition that negative functional values are set to zero. Here $f(t)$ is  the flux as a 
function of time $t$ in MJD, $A_0$ is the \textsl{amplitude} in units of (3-20 keV) PCA $\rm cts~s^{-1}$ per PCU, 
$A_1$=1.25, $B_1$=1.87, and $B_2$=0.80 are fixed functional parameters,
$C_1$ is the time reference in MJD, and $C_2$=$C_1$+8.4\,d (this offset represents the length of the 
standard  \textsl{Main-On}). The time of reference $C_1$ was fixed to the MJD of the turn-on 
that had been determined from the corresponding ASM light curve. 
The \textsl{amplitude} $A_{0}$ is the only free-fitting parameter. The maximum of the fitting function (on average found to be 
equal to $A_0$-5.4\,$\rm cts~s^{-1}$) is then taken as the \textsl{maximum flux} (in units of PCA $\rm cts~s^{-1}$)  for this cycle.
Dividing this  \textsl{maximum flux} by the corresponding mean flux of the complete light curve (containing all the
photons that are also used in generating the corresponding spectrum) leads to a scaling factor. Multiplying
the 3.5--60\,keV flux found from the spectra by this scaling factor then leads to the \textsl{maximum bolometric flux} 
(in units of $\rm erg~cm^{-2}~s^{-1}$). 
These bolometric flux values can now be compared to the \textsl{maximum ASM flux} 
values used in the earlier work by \citet{Staubert_etal_07}.

To test whether the above procedure in determining the maximum bolometric fluxes could
be influenced by systematic changes of the spectral shape over the 35\,d phase, we have investigated spectra
of cycle no. 323 which provides a good coverage of a complete \textsl{Main-On}. Systematic changes were only found
in one spectral parameter, the power-law index, with a magnitude of $\pm 3$\%, which corresponds to a negligible relative change
of the integrated flux of less than $5\times10^{-2}\%$.

The uncertainties in the bolometric flux values are dominated by the uncertainties in the scaling factors,
which in turn are dominated by systematic uncertainties in determining the maximum \textsl{Main-On} flux
in fitting the observed light curve by the common analytical function. These systematic uncertainties are estimated
to be 3\%,
the main contributions being the uncertainty in the turn-on time and the fluctuations in the shape of the modulation 
from cycle to cycle.

Table 2 compares the results of this re-analysis to those of the original 
analysis by \citet{Staubert_etal_07}. 

\section{Results and discussion} 

In Fig. \ref{compar} (left), we plot the newly determined cyclotron line energies against those from the original analysis. 
Overall, we find a very good agreement between the new values for E$_{\rm cyc}$ and those from the original analysis:
considering the respective uncertainties, the differences are smaller than 0.79 standard deviations for ten of the eleven 
values and  2.1 standard deviations for one value (cycle no. 323).

In Fig.~\ref{compar} (right), the 3.5--60\,keV maximum bolometric fluxes are plotted against the corresponding ASM fluxes 
of the original analysis. There is a good linear relationship between the two fluxes: the slope (taking the uncertainties
in both variables into account) is 0.88$\pm0.15$ ($\rm 10^{-9}~erg~cm^{-2}~s^{-1}$)/(ASM cts s$^{-1}$), and the Pearson correlation coefficient is 0.82, corresponding to a
chance probability of no correlation of P$<10^{-3}$. This demonstrates that the 
\textsl{maximum ASM flux} can be taken as a good measure of the \textsl{maximum bolometric X-ray flux} 
(and luminosity) of Her~X-1 during the respective 35\,d cycle. The variation in flux from one 35\,d cycle to the next is small, 
such that the maximum observed flux of a particular 35\,d \textsl{Main-On} can be considered a good measure of the luminosity 
of the source during this particular cycle. This is why the maximum ASM flux was used as a reference in the original 
analysis by \citet{Staubert_etal_07}. 

The final correlation between the cyclotron line energy and the X-ray flux is given in Fig.~\ref{correl} in two 
ways: we correlate E$_{\rm cyc}$ from the re-analysis with the \textsl{scaled 3.5--60\,keV maximum 
Main-On flux} (Fig.~\ref{correl}, left), as well as the \textsl{maximum ASM flux} (Fig.~\ref{correl}, right). 
The corresponding slopes of the linear fits to these data (taking the uncertainties of both variables into 
account) and the corresponding Pearson correlation coefficients \textsl{r} are: (i) for the Main-On flux presented in
Fig.~\ref{correl} (left) slope = 0.62$\pm0.19$ (keV)/($\rm 10^{-9}~erg~cm^{-2}~s^{-1}$) and \textsl{r} = 0.62 (P=$2\times10^{-2}$); (ii) for the maximum ASM flux presented in  Fig.~\ref{correl} (right) 
slope = 0.67$\pm0.14$ (keV)/(ASM cts s$^{-1}$) and \textsl{r} = 0.90 (P=$10^{-4}$).

The correlation seen in Fig.~\ref{correl} (left) is somewhat less convincing than that of Fig.~\ref{correl} (right)
(and that in the original analysis of \citealt{Staubert_etal_07}). We attribute this to the unavoidably larger
uncertainties associated mainly with the scaling of the bolometric flux measured for the individual spectrum to the 
flux that does describe the maximum flux of the particular 35\,d cycle. The originally used \textsl{maximum ASM fluxes} 
are, in contrast, simple direct measurements.

\section{Summary} 
We have re-analyzed observations of Her~X-1 in its Main-On state by \textsl{RXTE} between 1996 and 2005 with respect to its X-ray 
spectrum. Using data from both instruments (PCA and HEXTE) we performed a spectral analysis of observations of eleven Main-Ons
and determined the cyclotron line energy E$_{\rm cyc}$ and the 3.5--60\,keV flux for each of those Main-Ons. This observed flux was 
then scaled to a flux representing the \textsl{maximum flux} of the particular Main-On. 
We conclude that the \textsl{maximum ASM flux} used in the original analysis \citep{Staubert_etal_07} 
can really be taken as a measure of the luminosity of the source because it scales well with the  \textsl{maximum 5--60\,keV flux}
estimated through the spectral analysis. This is evident from Fig.~\ref{compar} (right) that shows 
there is a good linear relationship between the bolometric 3.5--60\,keV flux with the 2--10\,keV
flux measured by \textsl{RXTE}/ASM (both fluxes refer to the \textsl{maximum 35\,d Main-On flux}). 

The information contained in Figs. 2 and 3 (left and right) provides an internally consistent picture. We consider 
the combined evidence of all the correlations shown as proof of the correctness of the positive correlation between 
the cyclotron line energy and source luminosity as suggested in \citet{Staubert_etal_07}.
A direct comparison to the original analysis can be made by considering Fig.~\ref{correl} (right): the slope of 
the linear best fit is determined to (0.67$\pm0.14$) keV/($\rm cts~s^{-1}$), which is in good 
agreement with the (0.66$\pm0.10$) keV/($\rm cts~s^{-1}$) found in the original analysis. The final and confirmed result 
with respect to this correlation can then be stated as follows: \textsl{the value of the cyclotron line energy 
E$_{\rm cyc}$ increases by $\sim$7\% for a change in flux of a factor of two.}

In conclusion, we briefly describe the physical significance of the observed dependence - of either sign - 
of the cyclotron line energy with source luminosity. The \textsl{negative correlation} (decrease of  E$_{\rm cyc}$
with luminosity), which has been repeatedly observed for high luminosity transients such as V~0332+53 and 4U~0115+63 
\citep{Mihara_etal_98,Mowlavi_etal_06,Nakajima_etal_06,Tsygankov_etal_06}, has been interpreted in the following way: 
when the mass accretion rate (and hence the luminosity) increases, the height of the radiative shock above the 
neutron star surface increases, leading to a decrease in the effective magnetic field strength in the scattering region 
and therefore to a decrease in E$_{\rm cyc}$. This is in line  with theoretical considerations about accretion in the 
(locally) super-Eddington regime \citep{Burnard_etal_91}. In the sub-Eddington regime, however, believed to be
realized in Her~X-1, the deceleration of the accreted material is thought to be due to Coulomb drag and collective
plasma effects \citep{Nelson_etal_93}. In this case, under an increased accretion rate the atmosphere is compressed 
by the ram pressure of the infalling material and the scattering region moves closer to the neutron star surface. 
This is equivalent to an increase in effective field strength and an increase in E$_{\rm cyc}$, hence to a
\textsl{positive correlation} with luminosity \citep{Staubert_etal_07}. 

We would also like to add  that the recent analysis of \citet{Klochkov_etal_11} supports the physical correlation
between the cyclotron line energy and the luminosity of three sources: 4U~0115+63, V~0332+53,
and Her~X-1. In this analysis data from a short time interval were used, and spectra were generated by summing up 
photons belonging to individual pulses in selected ranges of pulse amplitude. In this \textsl{pulse-to-pulse variability} 
study, the variations in X-ray flux (source luminosity) occur on timescales comparable to the duration of the individual 
pulses (the period of rotation of the neutron star). The previously found correlations (based on flux variations on 
much longer timescales) between the cyclotron line energy and the X-ray flux are reproduced: that is a \textsl{negative} 
correlation for the super-Eddington transient sources 4U~0115+63 and V~0332+53, and a \textsl{positive} correlation 
for the sub-Eddington source Her~X-1.

\begin{acknowledgements} 
D.V. and coauthors thank DLR for financial support through grant 50 OR 0702
\end{acknowledgements}

\bibliographystyle{aa}
\bibliography{cyc_flux_eng_rev_RSt} 

\end{document}